\documentclass[twocolumn,showpacs,superscriptaddress,aps,pre]{revtex4-2}
\usepackage{bm}
\usepackage{amssymb,amsmath}
\usepackage{graphicx}
\usepackage{color}
\usepackage[colorlinks=true,urlcolor=blue,citecolor=blue]{hyperref}

\begin{document}
%%%%%%%%%%%%%%%%%%%%%%%%%%%%%%%%%%%%%%%%%%%%%%%%%%%%%%%%%%%%%%%%%%%%%%%%%%%%%%%%%%%%%%%%
% ABSTRACT / TITLE
%%%%%%%%%%%%%%%%%%%%%%%%%%%%%%%%%%%%%%%%%%%%%%%%%%%%%%%%%%%%%%%%%%%%%%%%%%%%%%%%%%%%%%%%
\title{Fractional Brownian motion with mean-density interaction:  a myopic self-avoiding fractional stochastic process}
\author{Jonathan House}
\affiliation{Department of Physics, Missouri University of Science and Technology, Rolla, MO 65409, USA}
\author{Rashad Bakhshizada}
\affiliation{Department of Physics, Missouri University of Science and Technology, Rolla, MO 65409, USA}
\author{Skirmantas Janu\v{s}onis}
\affiliation{Department of Psychological and Brain Sciences, University of California, Santa Barbara, CA 93106, USA}
\author{Ralf Metzler}
\affiliation{Institute of Physics and Astronomy, University of Potsdam, D-14476
Potsdam-Golm, Germany}

\author{Thomas Vojta}
\affiliation{Department of Physics, Missouri University of Science and Technology, Rolla, MO 65409, USA}

\begin{abstract}
Fractional Brownian motion is a Gaussian stochastic process with long-range correlations in time; it has been shown to be a useful model of anomalous diffusion. Here, we investigate the effects of mutual interactions in an ensemble of particles undergoing fractional Brownian motion. Specifically, we introduce a mean-density interaction in which each particle in the ensemble is coupled to the gradient of the total, time-integrated density produced by the entire ensemble. We report the results of extensive computer simulations for the mean-squared displacements and the probability densities of particles undergoing one-dimensional fractional Brownian motion with such a mean-density interaction. We find two qualitatively different regimes, depending on the anomalous diffusion exponent $\alpha$ characterizing the fractional Gaussian noise. The motion is governed by the interactions for $\alpha < 4/3$ whereas it is dominated by the fractional Gaussian noise for $\alpha > 4/3$.  We develop a scaling theory explaining our findings. We also discuss generalizations to higher space dimensions and nonlinear interactions, the relation of our process to the ``true'' or myopic self-avoiding walk,  as well as applications to the growth of strongly stochastic axons (e.g., serotonergic fibers) in vertebrate brains.
\end{abstract}

\date{\today}

\maketitle

%%%%%%%%%%%%%%%%%%%%%%%%%%%%%%%%%%%%%%%%%%%%%%%%%%%%%%%%%%%%%%%%%%%%%%%%%%%%%%%%%%%%%%%%
\section{Introduction}
\label{sec:intro}
%%%%%%%%%%%%%%%%%%%%%%%%%%%%%%%%%%%%%%%%%%%%%%%%%%%%%%%%%%%%%%%%%%%%%%%%%%%%%%%%%%%%%%%%

Diffusive transport is a widespread phenomenon that occurs in numerous physical, chemical, and biological systems. Its scientific investigation encompasses two centuries, ranging from Robert Brown's seminal experiment in 1827 \cite{Brown1828} to cutting-edge research today. The modern notion of diffusion is based on the groundbreaking discoveries of Einstein \cite{Einstein_book56}, Smoluchowski \cite{Smoluchowski18}, and Langevin \cite{Langevin08} according to which  normal diffusion results from a stochastic process that is local in both time and space, fulfilling three conditions: (i) individual particles are independent of each other; (ii) the process features a finite correlation time after which individual increments are statistically independent, and (iii) the displacements over a correlation time are symmetrically distributed in the positive and negative directions and feature a finite variance. If these conditions are fulfilled, the central limit theorem holds, yielding the celebrated linear dependence $\langle x^2 \rangle \sim t$ of the mean-squared displacement of the moving particle on the elapsed time $t$ \cite{Hughes95}.

Anomalous diffusion, i.e., random motion that does not obey the linear relation $\langle x^2 \rangle \sim t$, can occur in systems that violate at least one of the conditions listed above. Anomalous diffusion is instead characterized by the power law $\langle x^2 \rangle \sim t^\alpha$ where $\alpha$ is the  anomalous diffusion exponent.  For $\alpha < 1$, the motion is subdiffusive (i.e., $\langle  x^2 \rangle$ grows slower than $t$), whereas it is superdiffusive  for $\alpha > 1$ (i.e., $\langle x^2 \rangle$ grows faster than $t$). Both types of motion have been experimentally observed in numerous systems; and different mathematical models have been proposed to describe the resulting data (for reviews see, e.g., Refs.\ \cite{MetzlerKlafter00,HoeflingFranosch13,BressloffNewby13,MJCB14,MerozSokolov15,MetzlerJeonCherstvy16,Norregaardetal17}
and references therein).

For example, anomalous diffusion can be caused by slowly decaying long-range correlations between the increments (steps) of the stochastic process.
The paradigmatic mathematical model for this situation is fractional Brownian motion (FBM) \cite{Kolmogorov40,MandelbrotVanNess68}, a non-Markovian self-similar Gaussian stochastic process with stationary increments. Positive, persistent correlations between the increments lead to superdiffusion ($1 < \alpha < 2$), whereas negative, anti-persistent correlations produce subdiffusion ($0 < \alpha < 1$). For $\alpha=1$, FBM is identical to normal Brownian motion with uncorrelated increments. FBM has been successfully applied to model the dynamics in a wide variety of systems including diffusion inside biological cells \cite{SzymanskiWeiss09,MWBK09,WeberSpakowitzTheriot10,Jeonetal11,JMJM12,Tabeietal13}, the dynamics of polymers \cite{ChakravartiSebastian97,Panja10}, electronic network traffic \cite{MRRS02}, as well as fluctuations of financial markets \cite{ComteRenault98,RostekSchoebel13}.

Recently, reflected FBM \cite{WadaVojta18,Guggenbergeretal19,VHSJGM20} was employed to explain the inhomogeneous spatial distribution of serotonergic fibers (axons) in vertebrate brains \cite{JanusonisDetering19,JanusonisDeteringMetzlerVojta20,JanusonisHaimanMetzlerVojta23}. To this end, the set of serotonergic fibers is modeled as an ensemble of FBM trajectories that propagate inside the brain, starting from the cell bodies in the brainstem.  This model successfully reproduces key features of the inhomogeneous  fiber density distribution in the brain (whereas normal Brownian motion would tend to produce a homogeneous distribution).
So far, different fibers have been treated as independent in this approach. However, experimental evidence in mouse models suggests that the growth of serotonergic fibers is sensitive to the extracellular levels of serotonin (released by the fibers themselves) \cite{Daubertetal10,Migliarinietal13,VicenziFoaGasperini21}, and that active self-repulsion (as opposed to physical volume exclusion) contributes to the distribution of serotonergic fibers in the brain \cite{Chenetal17}. Specifically, a lack of serotonin synthesis in the developing brain increases serotonergic fiber densities in some forebrain regions \cite{Migliarinietal13}, and a pharmacologically-induced increase in brain serotonin levels (using fluoxetine, a widely prescribed antidepressant) results in a decrease in the serotonergic fiber densities in some of the same regions \cite{Nazzietal19,Nazzietal24}. These findings suggest that growing fibers may be sensitive to the local coarse-grained density of the entire fiber ensemble and be repulsed from regions where this density is high.

In this paper, we therefore introduce a stochastic process that models this idea. It consists of FBM with a ``mean-density'' interaction, i.e., an  interaction that couples each of the particles in a large ensemble of particles to the gradient of the total, time-integrated density of an entire ensemble. We then investigate, by means of large-scale computer simulations, the behavior of this stochastic process in one dimension. We find two qualitatively different regimes. If the anomalous diffusion exponent $\alpha$ of the underlying FBM is below 4/3, the motion is governed by the interactions whereas it is dominated by the fractional Gaussian noise for $\alpha > 4/3$.  To explain this interesting threshold behavior, we develop a one-parameter scaling theory.

Interestingly, our process can be understood as a generalization of the ``true'' or myopic self-avoiding random walk \cite{AmitParisiPeliti83} to fractional noise, and could therefore also be called ``myopic self-avoiding fractional Brownian motion''

Our paper is organized as follows. We define FBM, introduce the mean-density interaction, and discuss the details of our numerical approach  in Sec.\ \ref{sec:mod}.
In Sec.\ \ref{sec:results-msd},  we present the simulation results for the mean-squared displacement. The scaling theory is developed in Sec.\ \ref{sec:scaling}. In Sec.\ \ref{sec:results-prob}, we present simulation results for the instantaneous and integrated probability densities and compare them to the scaling theory predictions. Effects of fluctuations due to a finite particle number are addressed in Sec.\
\ref{sec:finite-N}.
We also consider generalizations to higher space dimensions and nonlinear interactions in Sec.\ \ref{sec:general}, and we conclude in Sec.\ \ref{sec:conclusions}.

%%%%%%%%%%%%%%%%%%%%%%%%%%%%%%%%%%%%%%%%%%%%%%%%%%%%%%%%%%%%%%%%%%%%%%%%%%%%%%%%%%%%%%%%
\section{Model}
\label{sec:mod}
%%%%%%%%%%%%%%%%%%%%%%%%%%%%%%%%%%%%%%%%%%%%%%%%%%%%%%%%%%%%%%%%%%%%%%%%%%%%%%%%%%%%%%%%
\subsection{Fractional Brownian motion}
\label{subsec:FBM}
%%%%%%%%%%%%%%%%%%%%%%%%%%%%%%%%%%%%%%%%%%%%%%%%%%%%%%%%%%%%%%%%%%%%%%%%%%%%%%%%%%%%%%%%

FBM can be defined as a continuous-time centered Gaussian stochastic process for the position $X$ of a particle that is located at the origin at time $t=0$.
In the absence of boundaries, the covariance function of the position at later times $s$ and $t$ is given by
\begin{equation}
\langle X(s) X(t) \rangle = K (s^\alpha - |s-t|^\alpha + t^\alpha)~,
\label{eq:FBM_cov}
\end{equation}
with $\alpha$ in the range $0 < \alpha < 2$. The constant $K$, of physical dimension length$^2$/time$^\alpha$, is the generalized diffusion   coefficient.
Setting $s=t$, this relation simplifies to $\langle X^2 \rangle = 2 K t^\alpha$ showing that FBM leads
to anomalous diffusion with anomalous diffusion exponent $\alpha$.
The corresponding probability density function of the position variable takes the Gaussian form
\begin{equation}
	P(X,t) = \frac{1}{\sqrt{4\pi K t^\alpha}} \exp{ \left( -\frac{X^2}{4 K t^\alpha} \right) }~.
\label{eq:FBM_P(x)_free}
\end{equation}

For computer simulations, it is convenient to work with a discrete-time version
of FBM \cite{Qian03}. We  discretize time by defining $x_n = X(t_n)$ with $t_n= \epsilon n$ where
$\epsilon$ is the time step and $n$ is an integer. The time evolution of the position $x_n$ takes the
form of a  random walk with identically Gaussian distributed but long-range
correlated steps, governed by the recursion relation
\begin{equation}
x_{n+1} = x_n + \xi_n~.
\label{eq:FBM_recursion}
\end{equation}
Here, the increments or steps $\xi_n$ constitute a discrete fractional Gaussian noise, i.e., a stationary
Gaussian process of zero mean, variance $\sigma^2 = 2 K \epsilon^\alpha$, and covariance function
\begin{equation}
C_n=\langle \xi_m \xi_{m+n} \rangle = \frac 1 2 \sigma^2 (|n+1|^\alpha - 2|n|^\alpha + |n-1|^\alpha)~.
\label{eq:FGN_cov}
\end{equation}
In the marginal case, $\alpha=1$, the covariance vanishes for all
$n\ne 0$, i.e., we recover normal Brownian motion. For $n\to \infty$, the covariance takes the
power-law form $\langle \xi_m \xi_{m+n} \rangle  \sim\alpha (\alpha-1) |n|^{-\gamma}$
with $\gamma=2-\alpha$.
The correlations are positive (persistent) for $\alpha>1$ and negative (anti-persistent)
for $\alpha < 1$.

To achieve the continuum limit, the standard deviation $\sigma$ of an individual step must be small compared to the
considered distances. Equivalently, the time step $\epsilon$ must be small compared to the total time $t$. The continuum
limit can thus be reached either by taking $\epsilon$ to zero at fixed $t$ or by taking $t$ to infinity at fixed $\epsilon$.
In this paper, we fix $\epsilon=1$ and consider the long-time limit $t \to \infty$.

%%%%%%%%%%%%%%%%%%%%%%%%%%%%%%%%%%%%%%%%%%%%%%%%%%%%%%%%%%%%%%%%%%%%%%%%%%%%%%%%%%%%%%%%
\subsection{Mean-density interaction}
\label{subsec:interaction}
%%%%%%%%%%%%%%%%%%%%%%%%%%%%%%%%%%%%%%%%%%%%%%%%%%%%%%%%%%%%%%%%%%%%%%%%%%%%%%%%%%%%%%%%

We now consider a large ensemble of $N$ particles, each performing an independent FBM process starting at time $t=0$.
In addition, the particles experience a generalized ``force'' that is proportional to the gradient of the total time-integrated density of the entire ensemble
since the starting time,
\begin{equation}
P_\mathrm{tot}(x, t_n) =  \sum_{j=1}^N  \sum_{m=1}^n \delta [x- x^{(j)}_m] ~,
\label{eq:Pc_discrete}
\end{equation}
where $x^{(j)}_m$ is the position of particle $j$ at time step $m$, and $\delta(x)$ denotes the Dirac delta function.

This is an appropriate choice, for example, for the application of the process to serotonergic fibers, as discussed in Sec.\ \ref{sec:intro}.  In this application, each growing fiber is represented by an FBM trajectory.  Fibers release serotonin along their entire length at a roughly constant rate (per time and per length of fiber).  Serotonin diffuses away and decays slowly, leading to a quasi-steady state in which the local serotonin density is approximately proportional to the density of fiber segments  $P_\mathrm{tot}(x, t_n)$  in a given region.  The growing fibers sense this density and are repulsed from high-density regions. Analogous arguments would apply to other applications in which the trajectories represent growing physical objects that persist in time.

We note that one can imagine other applications in which the influence of the density at earlier times decays with time, for example if
the interaction is mediated by a chemical with a finite life time that is released by the moving particle (rather than the entire trajectory). This situation can be modeled by introducing a memory kernel in Eq.\ (\ref{eq:Pc_discrete}). We will briefly consider this case in Appendix
\ref{app:decay}.

In the presence of a density dependent force, the recursion relation for the position of particle $j$,
\begin{equation}
x_{n+1}^{(j)}= x_{n}^{(j)} + \xi_{n}^{(j)}  + f[x_n^{(j)},t_n]~,
\label{eq:recursion_with_force}
\end{equation}
now contains two terms, the fractional Gaussian noise $\xi_{n}^{(j)}$ with covariance
\begin{equation}
\langle \xi_m^{(i)} \xi_{m+n}^{(j)} \rangle = C_n \delta_{ij}
\label{eq:FGN_cov_ensemble}
\end{equation}
(where $\delta_{ij}$ is the Kronecker $\delta$) and the force term
\begin{eqnarray}
f[x_n^{(j)},t_n] &=&  - \frac A N  \left .\frac {\partial} {\partial x} P_\mathrm{tot}(x,t_n) \right |_{x=x_n^{(j)}} \\
                       &=& - A \left .\frac {\partial} {\partial x} P_\mathrm{int}(x,t_n) \right |_{x=x_n^{(j)}} ~.
\label{eq:force}
\end{eqnarray}
Here, $P_\mathrm{int} = P_\mathrm{tot}/N$ is the mean integrated density of the ensemble. The factor $1/N$ in the relation between the total integrated density $P_\mathrm{tot}$ and the force is introduced in the spirit of mean-field theory to permit a well-defined thermodynamic limit $N\to \infty$. The parameter $A$ controls the character and strength of the interaction. For positive $A$, the particles are pushed away from regions of high density, whereas they are attracted to high-density regions for negative $A$. Note that the normalization of $P_\mathrm{int}$ is proportional to time
\begin{equation}
\int_{-\infty}^\infty dx P_\mathrm{int}(x,t_n) = n~,
\label{eq:Pc_norm}
\end{equation}
reflecting the growth of the trajectories with time.

In the application of FBM to the growth of serotonergic neurons in vertebrate brains discussed in Sec.\ \ref{sec:intro}, $P_\mathrm{tot}$ represents the total density of the growing set of serotonergic fibers. Assuming that the fibers are repulsed from regions of higher density, we are interested in positive $A$ in the following.

%%%%%%%%%%%%%%%%%%%%%%%%%%%%%%%%%%%%%%%%%%%%%%%%%%%%%%%%%%%%%%%%%%%%%%%%%%%%%%%%%%%%%%%%
\subsection{Simulation details}
\label{subsec:simdetails}
%%%%%%%%%%%%%%%%%%%%%%%%%%%%%%%%%%%%%%%%%%%%%%%%%%%%%%%%%%%%%%%%%%%%%%%%%%%%%%%%%%%%%%%%

We have performed computer simulations of discrete-time one-dimensional FBM with mean-density interaction for anomalous diffusion exponents $\alpha$ in the range
between 0.4 (in the subdiffusive regime) and 1.7 (deep in the superdiffusive regime).  We fix the time step at $\epsilon=1$ and set $K=1/2$, leading to a variance $\sigma^2=1$
of the individual steps. The particles start at the origin $x=0$ at $t=0$ and perform up to $2^{27} \approx 134$ million time steps.

The correlated Gaussian random numbers $\xi_n$  that represent the fractional noise for each particle are precalculated before the simulation by means of the Fourier-filtering method \cite{MHSS96}. This technique starts from a sequence of independent Gaussian random numbers $\chi_n$ of zero mean and unit variance (which we generate using the Box-Muller transformation with the LFSR113 random number generator proposed by L'Ecuyer \cite{Lecuyer99} as well as the 2005 version of Marsaglia's KISS \cite{Marsaglia05}).
The Fourier transform $\tilde \chi_\omega$ of these numbers is converted via ${\tilde{\xi}_\omega} = [\tilde C(\omega)]^{1/2} \tilde{\chi}_\omega$, using the
 Fourier transform  $\tilde C(\omega)$  of the covariance function (\ref{eq:FGN_cov}).
The inverse Fourier transformation of the ${\tilde{\xi}_\omega}$ yields the fractional Gaussian noise
\footnote{The Fourier filtering method uses a window size of $2 N_t$ where $N_t$ is the number of time steps. This ensures that the covariance of the resulting increments $\xi_n$ is given by (\ref{eq:FGN_cov}) and not distorted by the periodicity of the Fourier transformation.}.

To implement the mean-density interaction, we consider ensembles of up to $N=128$ particles. The mean integrated density $P_\mathrm{int}(x,t_n)$ is collected as a histogram
with a narrow bin width $\Delta x = 0.1$. To achieve a smooth mean integrated density for our moderately large ensemble sizes, we replace the $\delta$ functions in the
definition (\ref{eq:Pc_discrete}) of $P_\mathrm{int}(x,t_n)$ by Gaussians of variance 0.25.  We have confirmed that small changes of the histogram bin width and the variance of the smoothing Gaussian only lead to minuscule changes of the results
\footnote{If the $\delta$ functions in the definition (\ref{eq:Pc_discrete}) of $P_\mathrm{int}(x,t_n)$ are not smoothed, however, the histogram of the mean integrated density $P_\mathrm{int}(x,t_n)$ develops an unphysical oscillatory instability.}.
The gradient in the definition of the force (\ref{eq:force}) is computed via a simple two-point formula directly from the histogram. To ensure the robustness of the results, we have also performed test calculations using a higher-order (four-point) gradient formula as well as calculations that do not use a histogram at all but analytically compute the derivative of the sum of the accumulated Gaussians. The latter method is restricted to times $t \lessapprox 10^4$ because of the high numerical effort of keeping track of all the Gaussians. All gradient algorithms give the same results (within our error bars) for the available simulation times.

To further reduce the statistical errors, the results are averaged over up to 4096 independent ensembles, depending on the parameters

%%%%%%%%%%%%%%%%%%%%%%%%%%%%%%%%%%%%%%%%%%%%%%%%%%%%%%%%%%%%%%%%%%%%%%%%%%%%%%%%%%%%%%%%
\section{Results: Mean-squared displacement}
\label{sec:results-msd}
%%%%%%%%%%%%%%%%%%%%%%%%%%%%%%%%%%%%%%%%%%%%%%%%%%%%%%%%%%%%%%%%%%%%%%%%%%%%%%%%%%%%%%%%

\begin{figure}
\includegraphics[width=\columnwidth]{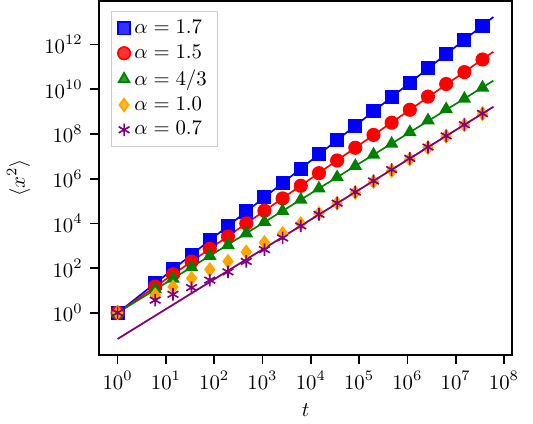}
\caption{Mean-squared displacement $\langle x^2 \rangle$ of FBM with mean-density interaction vs.\ time $t$ for interaction strength $A=1/40$ and several $\alpha$.
The data are averages over 16 ensembles of 128 random walkers each. The resulting statistical errors are much smaller than the symbol size. The solid lines are
power-law fits of the long-time behavior with $\langle x^2 \rangle = c\, t^{\bar\alpha}$.  They yield $\bar\alpha=\alpha$ for $\alpha\ge 4/3$ and $\bar\alpha=4/3$ for $\alpha < 4/3$, for details see text.}
\label{fig:MSD_alpha}
\end{figure}

We now turn to the results of our computer simulations.  Figure \ref{fig:MSD_alpha} presents the time evolution of the mean-squared displacement $\langle x^2 \rangle$
of several ensembles of random walkers performing FBM with mean-density interaction.
In all cases, the mean-squared displacement follows a power-law time dependence $\langle x^2 \rangle \sim t^{\bar\alpha}$ for sufficiently long times. Note that we need to distinguish the exponent $\alpha$ that parameterizes  the fractional Gaussian noise, as defined in Eq.\ (\ref{eq:FGN_cov}), from the exponent $\bar\alpha$ that characterizes the mean-squared displacement \emph{of the interacting system}.

A detailed analysis of the data in Fig.\ \ref{fig:MSD_alpha} reveals two different regimes. For $\alpha=1.7$, 1.5, and 4/3,
the mean-squared displacement features a power-law behavior over the entire time range. Fits with $\langle x^2 \rangle = c t^{\bar\alpha}$ where both $c$ and $\bar\alpha$
are fit parameters yield $\bar\alpha$ values very close to the corresponding  FBM value $\alpha$. In fact, the data can be fitted with high quality (reduced $\chi^2$ values below unity) with $\bar\alpha$ fixed at $\bar\alpha=\alpha$. The solid lines in Fig.\ \ref{fig:MSD_alpha} for $\alpha=1.7$, 1.5, and 4/3 show these fits.

The data for $\alpha=0.7$ and 1.0, in contrast, show a more complex behavior. At short times, the mean-squared displacement $\langle x^2 \rangle$ increases more slowly, as would be expected for (noninteracting) FBM with a lower $\alpha$. For longer times, $\langle x^2 \rangle$ crosses over to a faster power-law behavior that can be fitted very well (reduced $\chi^2$ values below unity) with $\langle x^2 \rangle \sim t^{\bar\alpha}$ with $\bar\alpha=4/3$ for both $\alpha=0.7$ and 1.0. The solid lines in Fig.\ \ref{fig:MSD_alpha} for $\alpha= 0.7$ and 1 correspond to such fits for times larger than $10^6$.   In fact, the mean-squared displacement curves for $\alpha=0.7$ and 1.0 are essentially indistinguishable for times beyond $10^5$. This suggests that the long-time behavior for these $\alpha$ values is dominated by interactions whereas the fractional Gaussian noise plays a subleading role.

\begin{figure}
\includegraphics[width=\columnwidth]{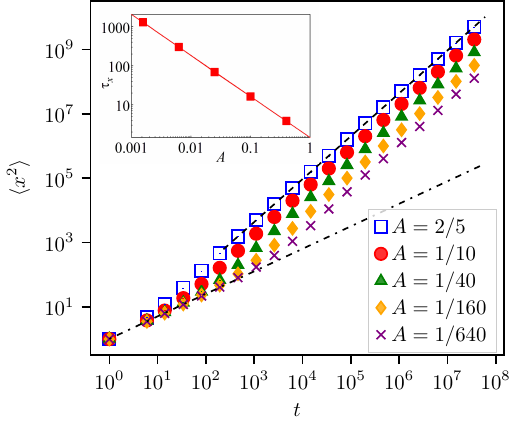}
\caption{Mean-squared displacement $\langle x^2 \rangle$ of FBM with mean-density interaction vs.\ time $t$ for $\alpha=0.7$ and several values of the interaction strength $A$. The data are averages over 60 ensembles of 128 random walkers each. The resulting statistical errors are much smaller than the symbol size. The dashed line represents a fit of the long-time behavior for $A=2/5$ with $\langle x^2 \rangle = c\, t^{4/3}$ whereas the dash-dotted line shows the FBM relation $\langle x^2 \rangle = \sigma^2 t^{0.7}$. Inset: crossover time $\tau_x$ from FBM to interaction-dominated behavior vs.\ interaction strength $A$. The solid line is a power law fit, yielding an exponent of $-1.05$. }
\label{fig:MSD_A}
\end{figure}

Further evidence for a crossover between FBM-like behavior at short times and interaction-dominated behavior at long times can be found in Fig.\ \ref{fig:MSD_A} which shows the mean-squared displacement at $\alpha=0.7$ for several different interaction strengths $A$.
At the earliest times, the mean-squared displacements are independent of $A$ and follow the FBM relation $\langle x^2 \rangle = \sigma^2 t^{0.7}$. After a crossover time $\tau_x$, the behavior of the mean-squared displacement changes to $\langle x^2 \rangle \sim t^{4/3}$ with an $A$-dependent prefactor. $\tau_x$ increases with decreasing interaction strength $A$, as can be seen in the inset of Fig.\  \ref{fig:MSD_A}.  We have observed an analogous behavior for $\alpha=1.0$. For $\alpha=1.5$, in contrast, the data for different $A$ are essentially indistinguishable because the force terms are negligible compared to the noise.

This crossover at $\tau_x$  for $\alpha < 4/3$ can be understood as follows. At short times, the integrated density $P_\mathrm{int}$ is small. The interaction terms (forces) (\ref{eq:force}) therefore do not yet play a role in the recursion relation (\ref{eq:recursion_with_force}), and the process behaves just like (non-interacting) FBM. As $P_\mathrm{int}$ increases with time, the interaction terms (\ref{eq:force}) also increase. At the crossover time $\tau_x$, their contribution to the displacement becomes comparable to that of the fractional Gaussian noise. The crossover time increases with decreasing interaction strength $A$ because, for smaller $A$, a larger integrated density $P_\mathrm{int}$ is required for the same generalized force $f$.  Beyond the crossover time, the process is interaction dominated, as discussed above.

\begin{figure}
\includegraphics[width=\columnwidth]{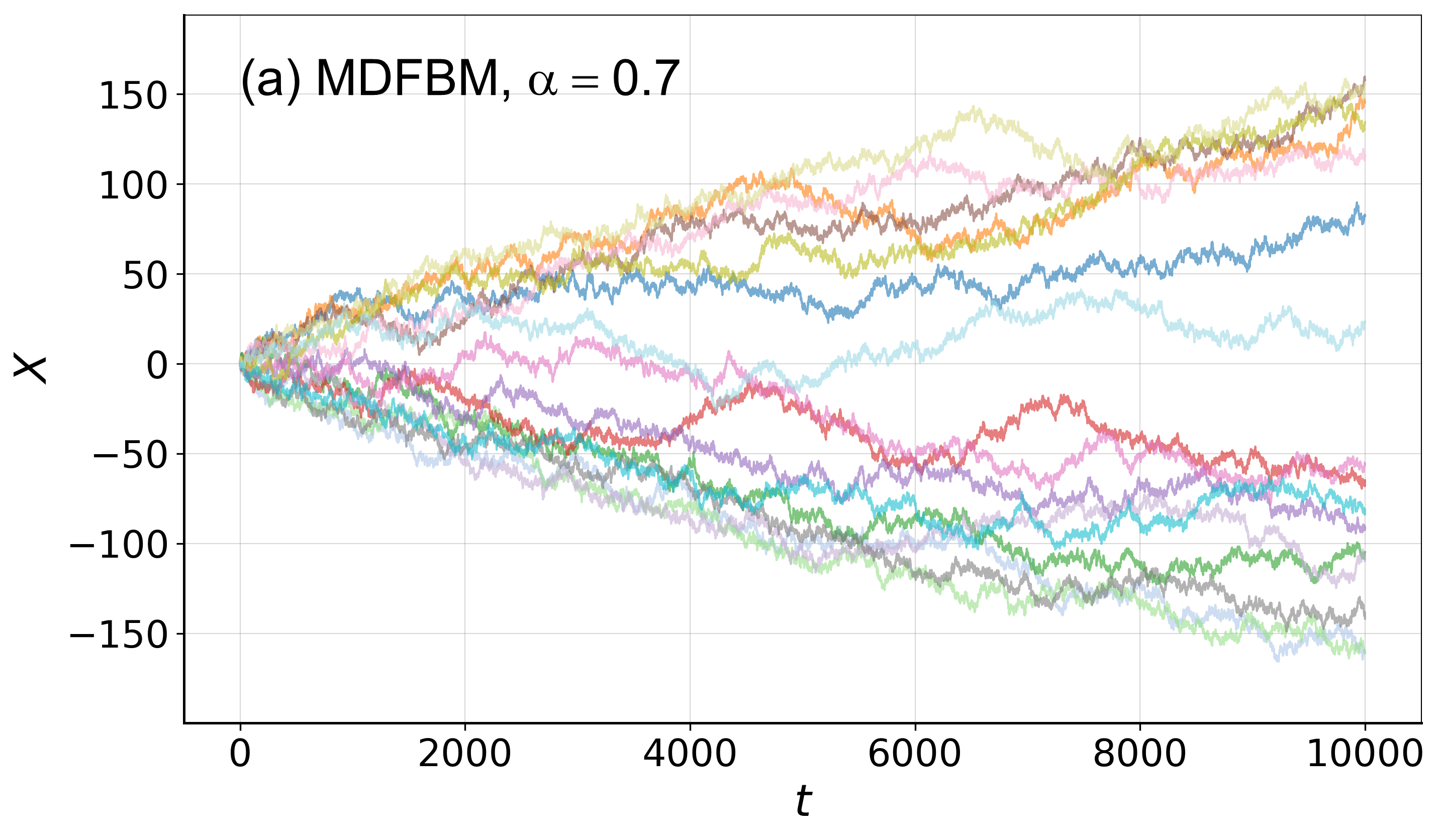}
\includegraphics[width=\columnwidth]{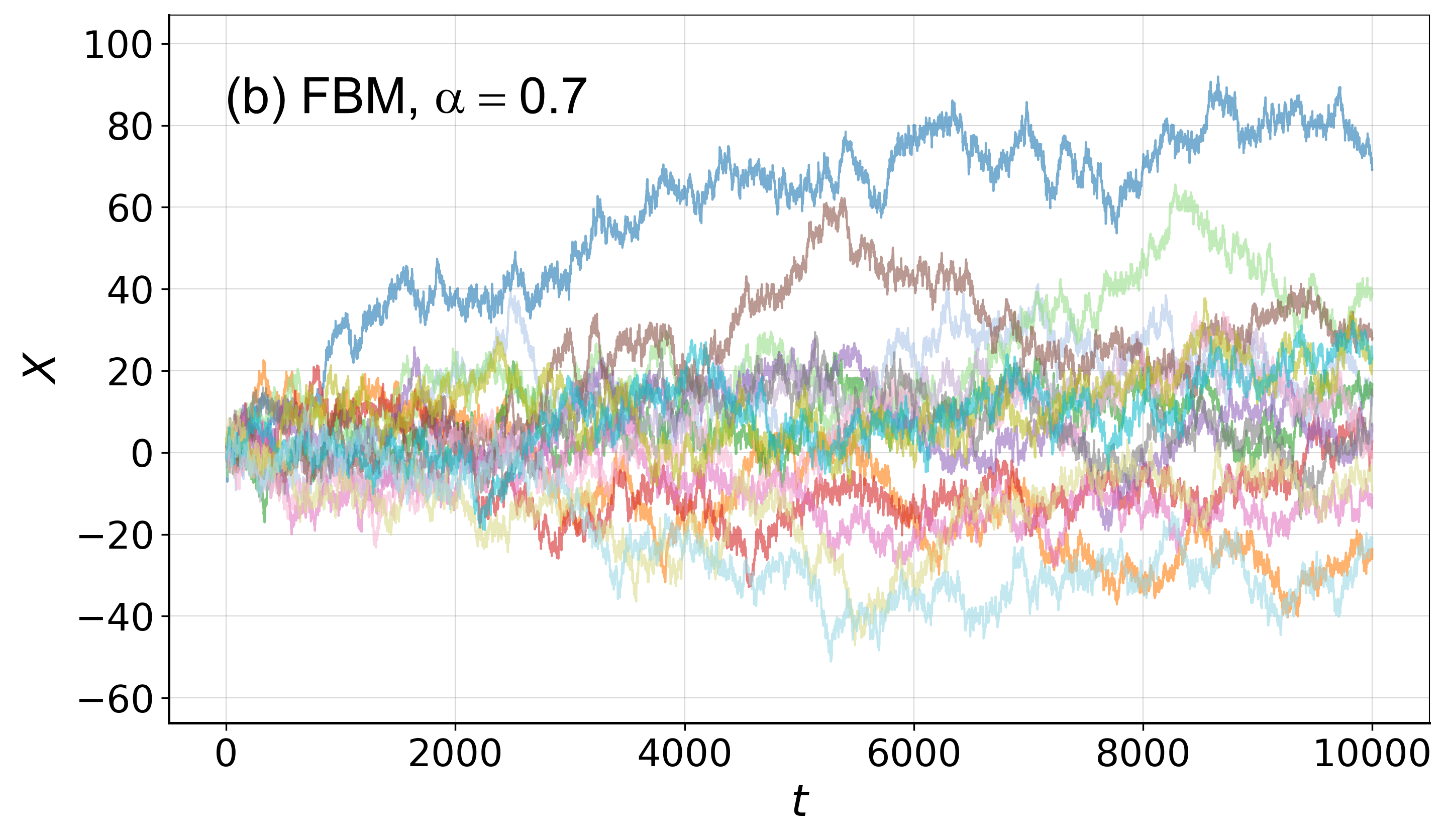}
\includegraphics[width=\columnwidth]{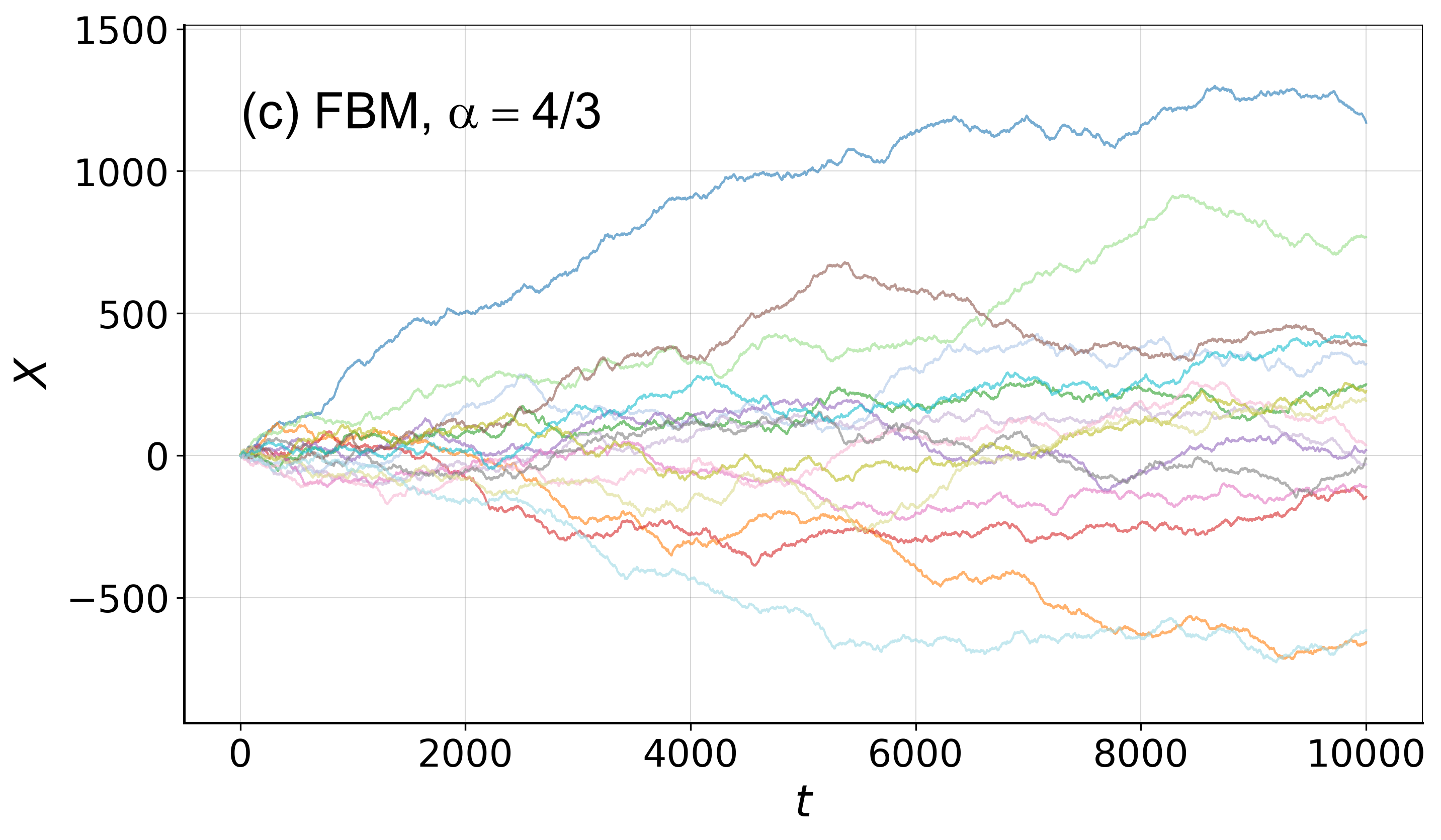}
\caption{Plots of 16 individual trajectories, randomly chosen from simulations with $N=128$ particles. (a) FBM with mean-density interaction for $\alpha=0.7$ and $A=1/40$.  (b) Non-interacting FBM, $\alpha=0.7$. (c) Non-interacting FBM, $\alpha=4/3$. (Note that FBM is self-similar; thus the appearance of the trajectories in panel (c) does not depend on the absolute scale.)  }
\label{fig:trajectories}
\end{figure}

It is also interesting to visualize individual trajectories. In the noise-dominated regime $\alpha > 4/3$, the trajectories closely resemble FBM trajectories because the interaction terms  (\ref{eq:force}) are small compared to the fractional Gaussian noise. We therefore focus on the interaction-dominated regime, $\alpha < 4/3$. Figure \ref{fig:trajectories} presents trajectories for $\alpha=0.7$ with and without the mean-density interaction. For comparison, it also shows FBM trajectories with $\alpha=4/3$.

The trajectories of non-interacting FBM with $\alpha=0.7$ shown in panel (b) are highly jittery (antipersistent) because of the negative correlations of subdiffusive ($\alpha < 1$) fractional Gaussian noise. This jittery motion is still visible in the presence of the mean-density interaction shown in panel (a). However, it is superposed onto a more regular motion away from the starting point (clearly visible for the trajectories furthest away from the origin (in the tails of the probability density). The comparison with panel (c) demonstrates that the trajectories of FBM with mean-density interaction do not resemble those of non-interacting FBM for $\alpha=4/3$ even though the mean-squared displacements of both processes evolve as $\langle x^2 \rangle \sim t^{4/3}$. The $\alpha=4/3$ FBM trajectories  are less jittery, even on short time scales because the fractional Gaussian noise for $\alpha >1$ is positively correlated \footnote{For non-interacting FBM in $d$ dimensions, the probability of trajectory crossings can be estimated from scaling arguments. During time $t$, a trajectory explores a length proportional to  $t^{\alpha/2}$, yielding a coincidence rate proportional to $t^{-d\alpha/2}$.  The number of crossings between times 0 and $t$ then reads $N(t) \sim \int_0^t d\tau \, \tau^{-d\alpha/2}$. Crossings proliferate for $d\alpha/2 <1$ which always holds in $d=1$. For the parameters relevant for the serotonergic fiber problem, $d\alpha/2>1$, and crossings are rare in free space. }.

%%%%%%%%%%%%%%%%%%%%%%%%%%%%%%%%%%%%%%%%%%%%%%%%%%%%%%%%%%%%%%%%%%%%%%%%%%%%%%%%%%%%%%%%
\section{Scaling theory}
\label{sec:scaling}
%%%%%%%%%%%%%%%%%%%%%%%%%%%%%%%%%%%%%%%%%%%%%%%%%%%%%%%%%%%%%%%%%%%%%%%%%%%%%%%%%%%%%%%%

In this section, we develop a scaling theory for FBM with mean-density interaction to explain the computer simulation results quantitatively. Consider an ensemble of $N$ random
walkers starting at the origin $x=0$ at time $t=0$. The scaling theory is based on the assumption that, for sufficiently long times, the  integrated distribution
$P_\mathrm{int}(x,t)$ approaches a universal functional form characterized by a single length scale $b(t)$ that increases with time $t$. This can be expressed via the scaling ansatz
\begin{equation}
P_\mathrm{int}(x,t) = \frac t {b(t)} Y\left [\frac{x} {b(t)} \right] ~.
\label{eq:scaling}
\end{equation}
The factor $t$ accounts for the fact that the space-integral (normalization) of the integrated density $P_\mathrm{int}(x,t)$ increases linearly with time. As a result, the scaling function $Y$ can be normalized to unity
\begin{equation}
\int_{-\infty}^{\infty} Y(y) dy =1~.
\label{eq:scaling_norm}
\end{equation}
Using this scaling form, the force term in the FBM recursion (\ref{eq:recursion_with_force}) can be expressed as
\begin{equation}
f(x,t) = -A \frac \partial {\partial x} P_\mathrm{int}(x,t) = - \frac{A t} {b^2(t)} Y' \left [\frac{x} {b(t)} \right]~.
\label{eq:scaling_force}
\end{equation}
Here $Y'$ denotes the derivative of the scaling function with respect to its argument.
Let us further assume that the length scale $b(t)$ increases according to the power law
\begin{equation}
b(t) \sim t^\delta ~,
\label{eq:b_vs_t}
\end{equation}
 with an unknown (positive) exponent $\delta$. Equation (\ref{eq:scaling_force}) then implies that the typical force varies with time as
\begin{equation}
f(x,t) \sim t^{1-2\delta}~.
\label{eq:f_vs_t}
\end{equation}
If the force term dominates the motion of the particles (compared to the fractional Gaussian noise), the leading behavior of the displacement is simply given by a time integral over the force. The typical
displacement is thus expected to behave as
\begin{equation}
x_\mathrm{typ}(t) \sim \int_0^t dt' \, f_\mathrm{typ}(t') \sim  t^{2-2\delta}~.
\label{eq:x_vs_t}
\end{equation}
For the theory to be self-consistent, the time dependence of $x_\mathrm{typ}$ needs to match the  assumed time dependence of the length scale $b(t)$,
\begin{equation}
t^\delta \propto t^{2-2\delta}~,
\label{eq:scaling_match}
\end{equation}
yielding $\delta=2/3$. In the force-dominated regime, the mean-squared displacement is therefore expected to behave as
\begin{equation}
\langle x^2 \rangle \sim b^2(t) \sim t^{4/3} ~.
\label{eq:MSD_scaling}
\end{equation}
To further check the self-consistency of the scaling theory, let us discuss what happens if the length scale $b(t)$ increases faster than $t^{2/3}$, as is expected to happen if the motion is driven by fractional Gaussian noise with an anomalous diffusion exponent $\alpha>4/3$. In this case, the displacement contribution (\ref{eq:x_vs_t}) resulting from integrating the forces would grow more slowly than $t^{2/3}$. This implies that the contribution of the forces to the displacement is subleading compared to the fractional
Gaussian noise.

If we assume, however, that the length scale $b(t)$ increases more slowly than $t^{2/3}$, the hypothetical contribution of the forces to the displacement would increase
faster than $t^{2/3}$, leading to a contradiction. The scaling theory therefore predicts that one-dimensional FBM with mean-density interaction in free space is dominated by the fractional Gaussian noise (and behaves like regular FBM) for $\alpha>4/3$  (i.e., $\gamma < 2/3)$, whereas it is interaction-dominated for $\alpha<4/3$  (i.e., $\gamma > 2/3$). This yields  the following mean-squared displacement behaviors,
\begin{equation}
\langle x^2 \rangle \sim \left\{ \begin{array}{ll}
t^{4/3} & \qquad \textrm{for }   \alpha < 4/3~,\\
t^{\alpha} = t^{2-\gamma}   & \qquad \textrm{for }  \alpha>4/3~.
\end{array}  \right. 
\label{eq:scaling_result}
\end{equation}
These predictions agree with the Monte Carlo results of Sec.\ \ref{sec:results-msd}.

The scaling theory also allows us to estimate the crossover time $\tau_x$ from the initial FBM behavior to the interaction-dominated long-time behavior.
According to Eq.\ (\ref{eq:scaling_force}), the forces behave as $f \sim A t^{1-\alpha}$ in the FBM regime, leading to a typical displacement $x_\mathrm{force} \sim A t^{2-\alpha}$. The crossover occurs when this force-induced displacement overcomes the FBM displacement.  This implies
$\sigma \tau_x^{\alpha/2} \sim A  \tau_x^{2-\alpha}$ or
\begin{equation}
\tau_x \sim \left( \frac A \sigma \right)^{2/(3\alpha-4)}~.
\label{eq:tau_x}
\end{equation}
This relation holds in the interaction-dominated regime,  $\alpha<4/3$. For the case of $\alpha=0.7$, the exponent in the power law (\ref{eq:tau_x}) evaluates to $-1.053$, in excellent agreement with the data in the inset of Fig.\ \ref{fig:MSD_A}.

%%%%%%%%%%%%%%%%%%%%%%%%%%%%%%%%%%%%%%%%%%%%%%%%%%%%%%%%%%%%%%%%%%%%%%%%%%%%%%%%%%%%%%%%
\section{Results: Probability densities}
\label{sec:results-prob}
%%%%%%%%%%%%%%%%%%%%%%%%%%%%%%%%%%%%%%%%%%%%%%%%%%%%%%%%%%%%%%%%%%%%%%%%%%%%%%%%%%%%%%%%

In this section, we present the Monte Carlo results for the mean integrated density $P_\mathrm{int}$, defined in eqs.\ (\ref{eq:Pc_discrete}) and (\ref{eq:force}) as
\begin{equation}
P_\mathrm{int}(x, t_n) = \frac 1 N  \sum_{j=1}^N  \sum_{m=1}^n \delta [x- x^{(j)}_m] ~.
\label{eq:Pint_discrete2}
\end{equation}
This not only provides additional insight into the behavior of FBM with mean-density interaction, but also allows us to test the assumptions underlying the scaling theory developed in Sec.\ \ref{sec:scaling}. In addition, we also analyze the instantaneous (fixed time) probability density of the diffusing particles,
\begin{equation}
P(x, t_n) = \frac 1 N  \sum_{j=1}^N \delta [x- x^{(j)}_n] ~.
\label{eq:Pi}
\end{equation}
In the application to serotonergic neurons, it represents the density of the active tips of the growing fibers.

For reference, we first consider the case of \emph{non-interacting} FBM.  The (instantaneous) probability density is given by Eq.\ (\ref{eq:FBM_P(x)_free}). It can be expressed in terms of the parameters of our discrete-time FBM version as
\begin{equation}
P(x,t_n) = \frac{1}{\sqrt{2\pi \sigma^2 t_n^\alpha}} \exp{ \left( -\frac{x^2}{2 \sigma^2 t_n^\alpha} \right) }~.
\label{eq:FBM_P(x)_discrete}
\end{equation}
The integrated density $P_\mathrm{int}(x,t_n)$ is obtained by summing this Gaussian over time steps 1 to $n$. In the continuum limit, the summation can be replaced by an integration, yielding
\begin{equation}
P_\mathrm{int}(x, t_n) = \frac{|x|^{2/\alpha-1} }{\alpha \pi^{1/2}(2\sigma^2)^{1/\alpha}}\, \Gamma\left( \frac 1 2 - \frac 1 \alpha, \frac{x^2}{2\sigma^2t_n^\alpha}\right) ~,
\label{eq:FBM_Pint}
\end{equation}
where $\Gamma$ is the upper incomplete $\Gamma$  function (for details, see Appendix \ref{app:idensity}). The function (\ref{eq:FBM_Pint}) has a maximum (with a cusp) at $x=0$ and a Gaussian tail (up to a power-law prefactor) for large $x$.

We now turn to our simulation results for the (instantaneous) probability density $P(x,t)$ and the time-integrated density $P_\mathrm{int}(x,t)$ for FBM with mean-density interaction. We have studied in detail two values of $\alpha$, one in the fractional-noise-dominated regime $\alpha > 4/3$ and one in the interaction-dominated regime $\alpha<4/3$.

\begin{figure}
\includegraphics[width=\columnwidth]{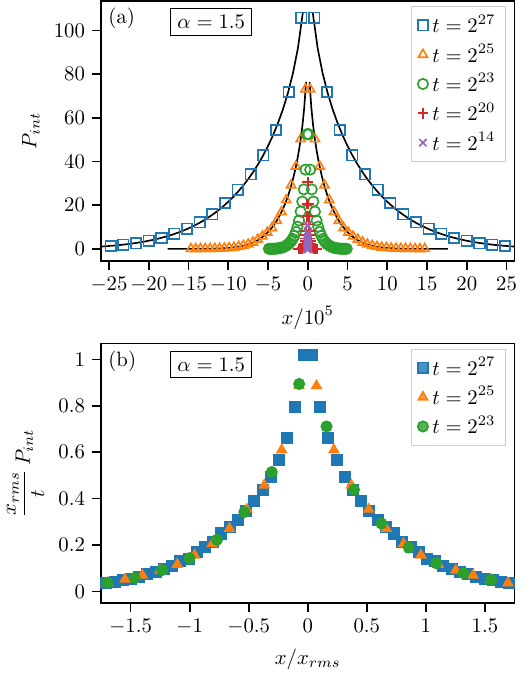}
\caption{(a) Integrated density $P_\mathrm{int}(x,t)$ for $\alpha=1.5$, $A=1/40$, and several $t$.  The data are averages over 120 ensembles of 64 random walkers each. To reduce the statistical noise in the figures, the histograms have been re-binned using 50 bins over the nonzero part of the histogram. The resulting statistical errors are much smaller than the symbol size. The solid lines shown for $t=2^{25}$ and $2^{27}$ correspond to the result (\ref {eq:FBM_Pint}) for noninteracting FBM.  (b) Scaled integrated density $x_\mathrm{rms} P_\mathrm{int}(x,t)/t$ vs.\ $x/x_\mathrm{rms}$ with $x_\mathrm{rms} =\sqrt{\langle x^2(t) \rangle}$.      }
\label{fig:fbm_int_dis}
\end{figure}

We start by discussing $\alpha=1.5$ in the noise-dominated regime. Figure \ref{fig:fbm_int_dis}(a) shows the integrated density for several different times.
As expected, $P_\mathrm{int}$ broadens with time, and its normalization increases, reflecting the growth of the trajectories with time.
Figure \ref{fig:fbm_int_dis}(a) also compares the simulation results for times $t=2^{25}$ and $2^{27}$ with the integrated density (\ref{eq:FBM_Pint}) of noninteracting FBM for the same $\alpha$. The agreement is nearly perfect and demonstrates that, for $\alpha=1.5$, the interaction does not affect the integrated density distribution at sufficiently long times. This agrees with the conclusion of the scaling theory of Sec.\ \ref{sec:scaling} which predicts that for $\alpha > 4/3$, the force terms in the recursion (\ref{eq:recursion_with_force}) become negligibly small compared to the fractional Gaussian noise for $t \to \infty$. Figure \ref{fig:fbm_int_dis}(b) shows that the integrated density fulfills the scaling form (\ref{eq:scaling}) with the root-mean-squared displacement $x_\mathrm{rms}(t) =\sqrt{\langle x^2(t) \rangle}$ playing the role of the length scale $b(t)$. This confirms the key assumption of the scaling theory.

\begin{figure}
\includegraphics[width=\columnwidth]{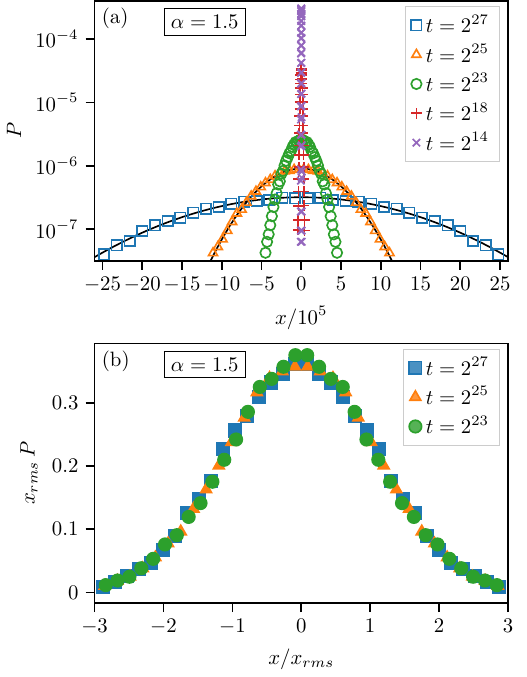}
\caption{(a)  (Instantaneous) probability density $P(x,t)$ for $\alpha=1.5$, $A=1/40$, and several $t$.  The data are averages over 120 ensembles of 64 random walkers each. To reduce the statistical noise in the figures, the histograms have been re-binned using 50 bins over the nonzero part of the histogram. The resulting statistical errors are about the symbol size. The solid lines shown for $t=2^{25}$ and $2^{27}$ correspond to the Gaussian distribution (\ref{eq:FBM_P(x)_discrete}) for noninteracting FBM. (b) Scaled probability density $x_\mathrm{rms} P(x,t)$ vs.\ $x/x_\mathrm{rms}$ with $x_\mathrm{rms} =\sqrt{\langle x^2(t) \rangle}$.    }
\label{fig:fbm_dis}
\end{figure}

In addition to the integrated density, we have also studied the (instantaneous) probability density $P(x,t)$. Simulation results for $\alpha =1.5$ are shown in Fig.\ \ref{fig:fbm_dis}(a).
In agreement with the notion that the interactions become negligible for long times, $P(x,t)$ agrees with the Gaussian distribution (\ref{eq:FBM_P(x)_discrete})
of (noninteracting) FBM.  Figure \ref{fig:fbm_dis}(b) confirms that the probability density fulfills the scaling form
\begin{equation}
P(x,t)= \frac 1{b(t)} Z \left[\frac x {b(t)} \right]~,
\label{eq:P_scaling}
\end{equation}
 with $b(t)=x_\mathrm{rms}(t) =\sqrt{\langle x^2(t) \rangle}$ and $Z$ being a dimensionless scaling function. Of course, for noninteracting FBM, this follows directly from Eq.\ (\ref{eq:FBM_P(x)_discrete}).

After having discussed  the fractional-noise-dominated regime $\alpha > 4/3$, we now consider the interaction-dominated regime $\alpha < 4/3$. This regime is arguably more interesting, because we expect the behavior of our process to differ qualitatively from that of noninteracting FBM. We have performed extensive simulations for $\alpha=0.7$ and 1.0 in the interaction-dominated regime. In the following, we discuss the case $\alpha=1.0$ as an example.

\begin{figure}
\includegraphics[width=\columnwidth]{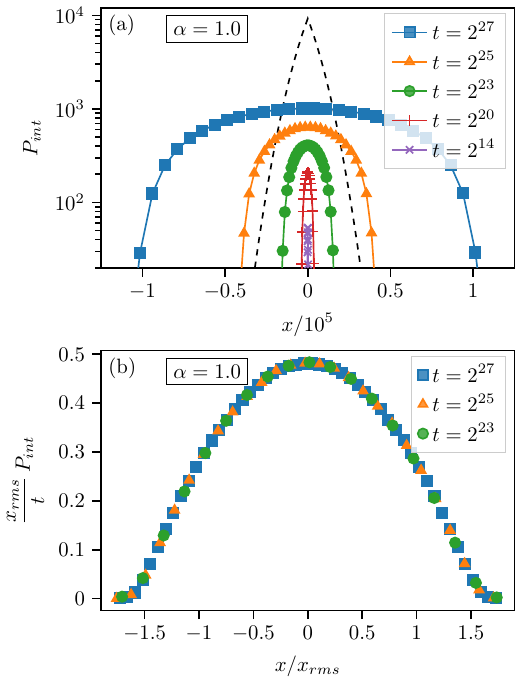}
\caption{(a) Integrated density $P_\mathrm{int}(x,t)$ for $\alpha=1.0$, $A=1/40$, and several $t$.  The data are averages over 120 ensembles of 64 random walkers each. To reduce the statistical noise in the figures, the histograms have been re-binned using 30 bins over the nonzero part of the histogram. The resulting statistical errors are much smaller than the symbol size. For comparison, the dashed line shows the result (\ref {eq:FBM_Pint}) for noninteracting FBM at time $t=2^{27}$.  (b) Scaled integrated density $x_\mathrm{rms} P_\mathrm{int}(x,t)/t$ vs.\ $x/x_\mathrm{rms}$ with $x_\mathrm{rms} =\sqrt{\langle x^2(t) \rangle}$.  }
\label{fig:force_int_dis}
\end{figure}

Figure \ref{fig:force_int_dis}(a) shows the time-integrated density $P_\mathrm{int}(x,t)$ for $\alpha=1.0$ and several values of the time $t$.
$P_\mathrm{int}$ broadens with time, and its normalization increases, as expected. The figure also presents (as a dashed line) the integrated density (\ref{eq:FBM_Pint}) of noninteracting FBM for the same $\alpha=1$ at time $t=2^{27}$. The data clearly show that the interacting integrated density is much broader than that of noninteracting FBM, and it has a different shape (in particular, no cusp at $x=0$). This agrees with the notion that, for $\alpha<4/3$, the interactions dominate the time evolution and lead to a more rapid expansion of the particle ``cloud'' than the fractional Gaussian noise would. Nonetheless, the integrated density fulfills the scaling form (\ref{eq:scaling}) with $b(t)=x_\mathrm{rms}(t) =\sqrt{\langle x^2(t) \rangle}$, as is demonstrated in Fig.\ \ref{fig:force_int_dis}(b). This confirms that the key assumption of the scaling theory holds not just in the fractional-noise-dominated regime but also in the interaction-dominated regime.

\begin{figure}
\includegraphics[width=\columnwidth]{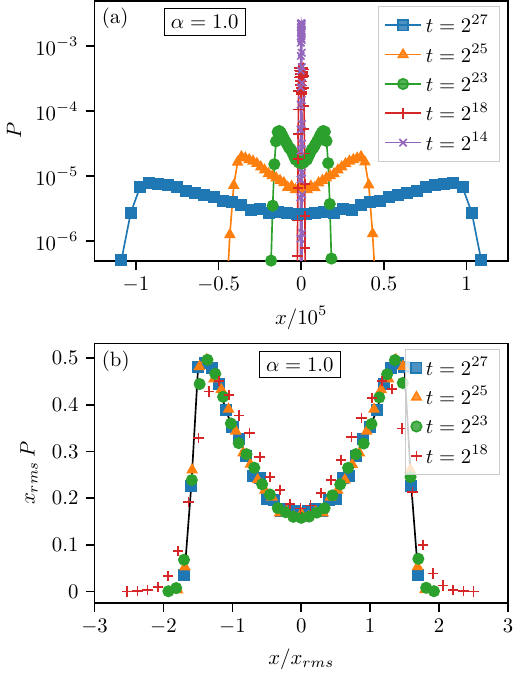}
\caption{(a)  (Instantaneous) probability density $P(x,t)$ for $\alpha=1.0$, $A=1/40$, and several $t$.  The data are averages over 120 ensembles of 64 random walkers each. To reduce the statistical noise in the figures, the histograms have been re-binned using 40 bins over the nonzero part of the histogram. The resulting statistical errors are about the symbol size. (b) Scaled probability density $x_\mathrm{rms} P(x,t)$ vs.\ $x/x_\mathrm{rms}$ with $x_\mathrm{rms} =\sqrt{\langle x^2(t) \rangle}$.  }
\label{fig:force_dis}
\end{figure}

Simulation results for the (instantaneous) probability density $P(x,t)$ for $\alpha=1.0$ are shown in Fig.\ \ref{fig:force_dis}(a).
The figure demonstrates that the probability density in the interaction-dominated regime differs significantly from that of FBM and is highly non-Gaussian. Interestingly, the maximum of $P(x,t)$ is not at the center $x=0$. Instead, there are two symmetric maxima after which $P(x,t)$
rapidly drops to zero. This can be understood as follows. In the interaction-dominated regime, the force terms in the recursion (\ref{eq:recursion_with_force}) push the particles strongly away from the center region where the integrated density (i.e., the density of the entire ensemble of trajectories) accumulated during previous time steps. At any given time, the ``active'' particles (i.e., the tips of the trajectories) are therefore concentrated near the boundary of the integrated density. For example, Fig.\ \ref{fig:force_int_dis}(a) shows that the integrated density at $t=2^{27}$ roughly extends to $x=\pm 10^5$. Correspondingly, the maxima of the instantaneous probability density in Fig.\ \ref{fig:force_dis}(a) for $t=2^{27}$ are at positions $x \approx \pm 10^5$.

Despite its highly non-Gaussian form, the probability density in the interaction-dominated regime fulfills the scaling form (\ref{eq:P_scaling}) with  $b(t)=x_\mathrm{rms}(t) =\sqrt{\langle x^2(t) \rangle}$, as can be seen in Fig.\ \ref{fig:force_dis}(b). The small deviations from perfect scaling collapse for the shortest time in the figure can be attributed to finite-time effects. Specifically, the force terms do not completely dominate at $t=2^{18}$, and the fractional Gaussian noise produces the small tails at large $|x|$.

We note that similar bimodal probability densities are observed, for different physical reasons, in L\'evy walks, certain heterogeneous diffusion processes, fractional wave equations, the end-to-end distance of semi-flexible polymers, as well as regular but confined FBM.

%%%%%%%%%%%%%%%%%%%%%%%%%%%%%%%%%%%%%%%%%%%%%%%%%%%%%%%%%%%%%%%%%%%%%%%%%%%%%%%%%%%%%%%%
\section{Finite-$N$ fluctuations}
\label{sec:finite-N}
%%%%%%%%%%%%%%%%%%%%%%%%%%%%%%%%%%%%%%%%%%%%%%%%%%%%%%%%%%%%%%%%%%%%%%%%%%%%%%%%%%%%%%%%

So far, we have considered large particle numbers ($N \gg 1$). In the limit $N\to \infty$, the integrated density of a given ensemble of $N$ particles  becomes identical to its average over all noise realizations, i.e., finite-$N$ fluctuations of $P_\mathrm{int}$ and the resulting forces are suppressed. In this section, we study the effects of a finite particle number (ensemble size) $N$.  We focus on the interaction-dominated regime $\alpha < 4/3$ because the forces do not affect the asymptotic behavior in the noise dominated regime $\alpha > 4/3$.

\begin{figure}
\includegraphics[width=\columnwidth]{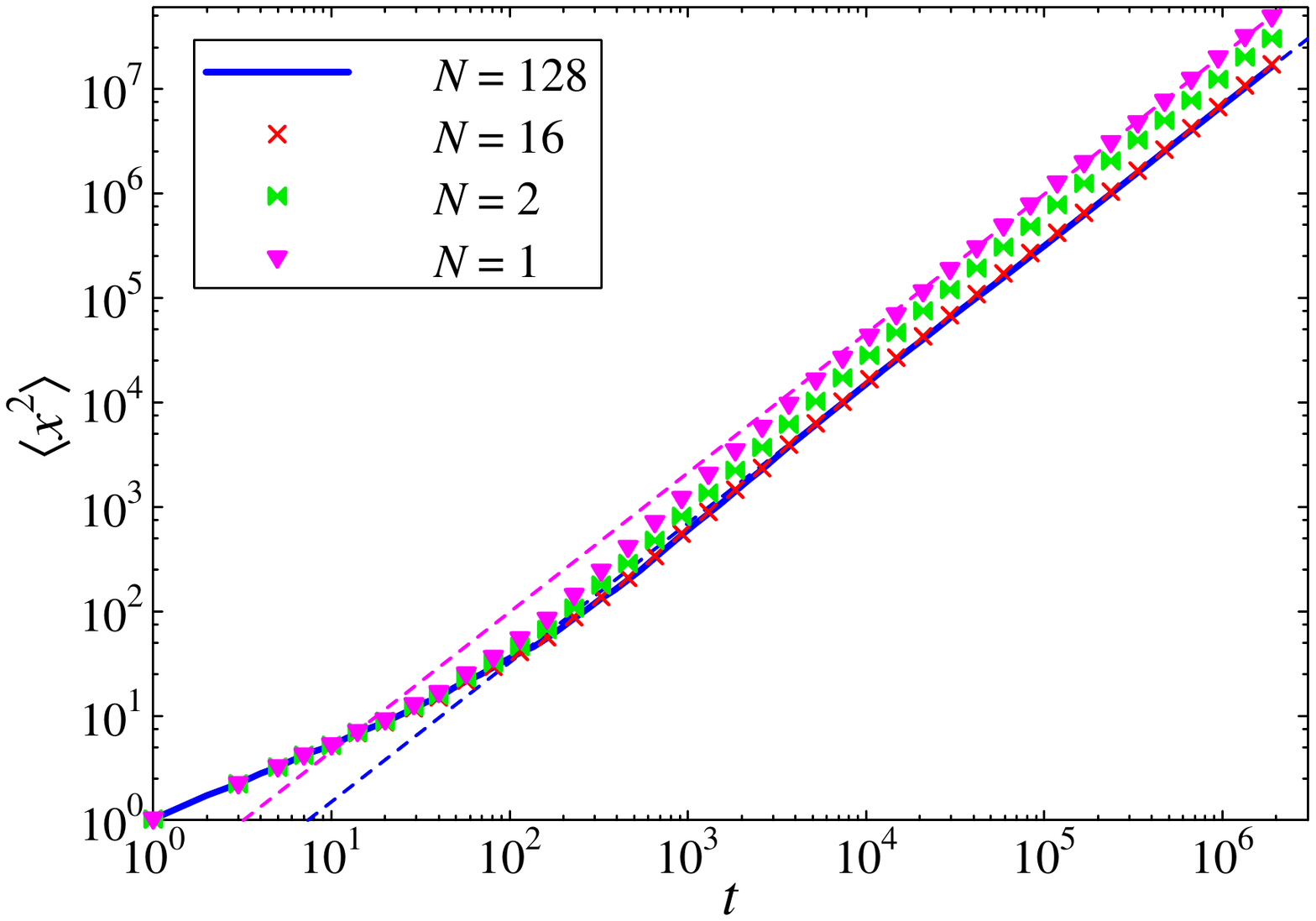}
\caption{Mean-squared displacement $\langle x^2 \rangle$ of FBM with mean-density interaction vs.\ time $t$ for $\alpha=0.7$, interaction strength $A=1/40$ and different particle numbers $N$. The data are averages over 2048 ensembles ($N=1$), 1024 ensembles ($N=2$), 128 ensembles ($N=16$), and 16 ensembles ($N=128$). The dashed  lines are fits of the long-time behavior for $N=1$ and $N=128$ with $\langle x^2 \rangle = c\, t^{4/3}$. }
\label{fig:MSD_N}
\end{figure}

Figure \ref{fig:MSD_N} presents the mean-squared displacement for $\alpha=0.7$ for particle numbers $N=1, 2, 16$, and 128.
The mean-squared displacement curve for $N=16$ is essentially indistinguishable from that for $N=128$. The same holds for $N=32$, and 64 (not shown in the figure for clarity). We conclude that the results for $N \ge 16$  represent the infinite-ensemble limit $N\to \infty$. The data for $N < 16$ show some deviations from the $N\to \infty$ behavior.  However, all curves (including the ones for $N<16$) asymptotically follow the same power law $\langle x^2 \rangle \sim t^{4/3}$. This indicates that the process belongs to the same universality class for all $N$ including $N=1$.

\begin{figure}
\includegraphics[width=\columnwidth]{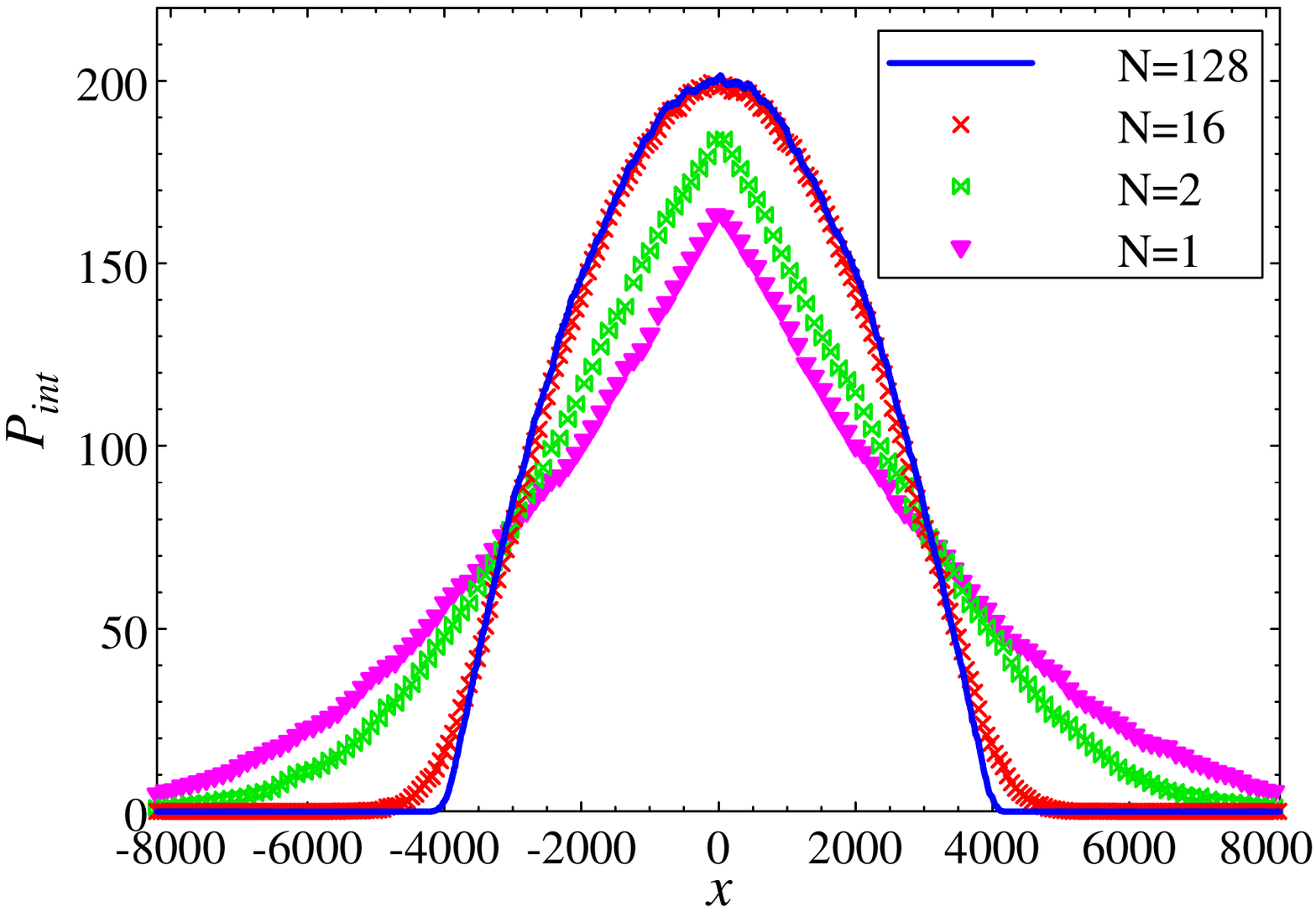}
\caption{Integrated density $P_\mathrm{int}(x,t)$ at time $t=2^{20}$ for $\alpha=0.7$, interaction strength $A=1/40$ and different particle numbers $N$. The data are averages over 4096 ensembles ($N=1$), 2048 ensembles ($N=2$), 256 ensembles ($N=16$), and 32 ensembles ($N=128$).}
\label{fig:Pint_N}
\end{figure}

This can be further illustrated by studying the integrated density $P_\mathrm{int}$ for different $N$. Figure \ref{fig:Pint_N} presents the integrated density at time $t=2^{20} \approx 10^6$ for $\alpha=0.7$ and particle numbers $N=1, 2, 16$, and 128.
(As can be seen in Fig.\ \ref{fig:MSD_N}, the system has reached the asymptotic regime for all $N$ at time $t=10^6$.)
The densities for $N=16$ and $N=128$ are almost indistinguishable and represent the infinite-particle-number limit. For smaller $N$,
the density broadens, in agreement with the observation that the mean-squared displacement in Fig.\ \ref{fig:MSD_N} increases with decreasing $N$ at fixed $t$.
We have confirmed that the density $P_\mathrm{int}$ fulfills the scaling form (\ref{eq:scaling}) for all $N$ with $b(t) \sim t^{2/3}$,
providing further evidence that the system belongs to the same universality class for all $N$.
However, the shape of the scaling function $Y$ is $N$-dependent.  Analogous behavior is expected for all $\alpha$ in the interaction-dominated regime $\alpha < 4/3$.

%%%%%%%%%%%%%%%%%%%%%%%%%%%%%%%%%%%%%%%%%%%%%%%%%%%%%%%%%%%%%%%%%%%%%%%%%%%%%%%%%%%%%%%%
\section{Generalizations}
\label{sec:general}
%%%%%%%%%%%%%%%%%%%%%%%%%%%%%%%%%%%%%%%%%%%%%%%%%%%%%%%%%%%%%%%%%%%%%%%%%%%%%%%%%%%%%%%%

In the preceding sections, we have analyzed motion in one space dimension under the influence of a force that is proportional to the gradient of the integrated density. It is interesting
to ask how the system behaves in higher dimensions and for other functional forms of the density-dependent force. The scaling theory of Sec.\ \ref{sec:scaling} is easily
generalized to $d$ space dimensions and forces that behave as the $\lambda$-th power of the gradient of $P_\mathrm{int}(\mathbf{x},t_n)$.

In $d$ dimensions, the scaling form (\ref{eq:scaling}) of the integrated mean density generalizes to
\begin{equation}
P_\mathrm{int}(\mathbf{x},t) = \frac t {b^d(t)} Y\left [\frac{\mathbf{x}} {b(t)} \right] ~.
\label{eq:scaling_d}
\end{equation}
If we again assume that the length scale $b(t)$ increases as $t^\delta$ with unknown $\delta$, the force term behaves as
\begin{equation}
|\mathbf f(\mathbf{x},t)| = \left | A \frac \partial {\partial \mathbf{x}} P_\mathrm{int}(\mathbf{x},t) \right |^\lambda  \sim t^{ \lambda -(d+1)\delta\lambda} ~.
\label{eq:scaling_force_d}
\end{equation}
In the force-dominated regime, the typical displacement is obtained by integrating this force over time. It is thus expected to behave as
$x_\mathrm{typ} \sim t^{ 1+\lambda -(d+1)\delta\lambda}$. Self-consistency with the assumption $b(t) \sim t^\delta$ requires
$1+\lambda -(d+1)\delta\lambda = \delta$. Solving for the value of $\delta$ yields
\begin{equation}
\delta = \frac {1+\lambda} {1+(d+1)\lambda}~.
\label{eq:delta_general}
\end{equation}
For $d=\lambda=1$, we recover the result of Sec.\  \ref{sec:scaling}, $\delta=2/3$.

Repeating the arguments at the end of Sec.\  \ref{sec:scaling}, we conclude that the motion of FBM with mean-density interaction will be interaction dominated if the
FBM anomalous diffusion exponent $\alpha$ is smaller than $2\delta$. In this case the mean-squared displacement is expected to behave as $\langle \mathbf{x}^2 \rangle \sim t^{2\delta}$.
If, however, $\alpha>2\delta$, then the motion will be dominated by the fractional Gaussian noise leading to  $\langle \mathbf{x}^2 \rangle \sim t^\alpha$.
Equation (\ref{eq:delta_general}) shows that $\delta$ decreases with increasing space dimensionality $d$. This implies that the marginal value of $\alpha$, below which the interactions dominate over the noise, decreases with increasing $d$. The result that the interaction effects are strongest in one dimension and decrease with increasing $d$ is perhaps not unexpected as crowding is more easily achieved in lower dimensions.

Let us specifically consider the cases $d=2$ and $d=3$ for linear gradient forces, $\lambda=1$. In two dimensions, Eq.\ (\ref{eq:delta_general}) reduces to $\delta= 1/2$. This implies that $\alpha=2\delta=1$ is the marginal value of $\alpha$ that separates the noise-dominated and interaction-dominated regimes in two dimensions. We note that this agrees with the finding that $d=2$ is the critical dimension of the so-called  ``true'' or myopic self-avoiding random walk \cite{AmitParisiPeliti83} discussed in Sec.\  \ref{sec:conclusions}.
Analogously, for $d=3$, we find $\delta=2/5$ from Eq.\ (\ref{eq:delta_general}), implying a critical $\alpha=0.8$.

Computer simulations of FBM with mean-density interaction in higher space dimensions require a significantly larger numerical effort. For this reason, a numerical test of the generalized scaling theory is relegated to future work.

%%%%%%%%%%%%%%%%%%%%%%%%%%%%%%%%%%%%%%%%%%%%%%%%%%%%%%%%%%%%%%%%%%%%%%%%%%%%%%%%%%%%%%%%
\section{Conclusions}
\label{sec:conclusions}
%%%%%%%%%%%%%%%%%%%%%%%%%%%%%%%%%%%%%%%%%%%%%%%%%%%%%%%%%%%%%%%%%%%%%%%%%%%%%%%%%%%%%%%%

In this paper, we have introduced FBM with mean-density interaction, a process in which each particle of an ensemble evolves under the influence of both fractional Gaussian noise and a force proportional to the gradient of the time-integrated density of the entire ensemble. This work was motivated by the recent application of (reflected) FBM to the growth of serotonergic fibers in vertebrate brains  \cite{JanusonisDetering19,JanusonisDeteringMetzlerVojta20,JanusonisHaimanMetzlerVojta23}. However, we believe our model to be applicable to a much broader class of anomalous diffusion processes in which the particles interact with a (coarse-grained) density of the resulting trajectories.

The present paper has focused on exploring the basic properties of this stochastic process in one space dimension. Employing large-scale computer simulations as well as a one-parameter scaling theory, we have found that the behavior of unbounded  FBM with mean-density interaction falls into one of two regimes, depending on the value of the exponent $\alpha$ characterizing the fractional Gaussian noise. For $\alpha > 4/3$,  the long-time behavior is governed by the fractional Gaussian noise, and the force terms become negligibly small. Consequently, in this regime, the mean-squared displacement and the probability density agree with the corresponding quantities of noninteracting FBM for sufficiently long times.   For $\alpha < 4/3$, in contrast, the long-time behavior of the model is dominated by the interactions, and the fractional Gaussian noise only makes subleading contributions. As a result, the mean-squared displacement grows like $t^{4/3}$ for all $\alpha \le 4/3$ and the probability density becomes highly non-Gaussian.

The stochastic process introduced in this paper is related to the so-called ``true'' or myopic self-avoiding random walk that was introduced into the physics literature by Amit, Parisi, and Peliti \cite{AmitParisiPeliti83} and further studied in Refs.\ \cite{Pietronero83,PelitiPietronero87,Lawler91}. The myopic self-avoiding random walk  is defined as the problem of a particle that performs (uncorrelated) random steps on a lattice but tries to avoid already visited sites. (Note that this process is different from the usual self-avoiding walk and belongs to a different universality class \cite{AmitParisiPeliti83}.)  Pietronero \cite{Pietronero83} showed that the mean-squared displacement of the myopic self-avoiding random walk in one dimension behaves as $\langle x^2 \rangle \sim t^{4/3}$ which agrees with our result for uncorrelated noise, $\alpha=1$.  In general dimension $d$, Pietronero found $\langle x^2 \rangle \sim t^{4/(d+2)}$ for $d<2$ and  $\langle x^2 \rangle \sim t$ for $d\ge 2$. This agrees with the results of our generalized scaling theory in Sec.\ \ref{sec:general} for $\alpha=1$ and $\lambda=1$. We thus conclude that \emph{normal} Brownian motion with mean-density interaction belongs to the same universality class as the myopic self-avoiding random walk.  Our process  constitutes a generalization of this universality class to fractional noise, and could therefore also be called ``myopic self-avoiding fractional Brownian motion'' \footnote{Strictly, the myopic self-avoiding random walk is related to the $N=1$ version of our process at $\alpha=1$. However, as was shown in Sec.\ \ref{sec:finite-N}, our process belongs to the same universality class for all $N$.}.

Our work suggests many interesting applications and extensions that may stimulate further research. These include the questions of higher space dimensions and nonlinear forces that we have already touched upon in Sec.\  \ref{sec:general}. It is also interesting to study what happens in applications in which the particles are attracted rather than repulsed by regions of high (integrated) density.

One important application is the problem that motivated the present study, viz.,  the system of serotonergic fibers in vertebrate brains. This application not only requires a generalization to three dimensions, it also implies simulations in a complex geometry. We expect a nontrivial interplay between the mean-density interaction and the reflecting walls that confine the process in the brain shape (or some other finite or semi-finite region of space). Specifically, we expect that the interaction will cut off the unphysical divergence of the density close to the wall observed for (non-interacting) superdiffusive FBM, potentially providing an improved description of the fiber system.  Due to the high dimensionality and complex shape, these simulations are expected to require a huge numerical effort and thus remain a task for future.

Our paper may also have interesting applications in ecology. Many animals leave traces (e.g., pheromones) along their paths which then affect the behavior of other animals. This is an example of stigmergy,  a mechanism of indirect interaction between agents via their environment \cite{Grasse59,TheraulazBonabeau99,Heylighen16a,Heylighen16b}.  Our stochastic process may provide a basis for modeling  stigmergic phenomena. Environmental traces often decay with time, this can be accounted for via a decaying memory kernel in the interaction, as discussed in  Appendix \ref{app:decay}.

%%%%%%%%%%%%%%%%%%%%%%%%%%%%%%%%%%%%%%%%%%%%%%%%%%%%%%%%%%%%%%%%%%%%%%%%%%%%%%%
\acknowledgments
%%%%%%%%%%%%%%%%%%%%%%%%%%%%%%%%%%%%%%%%%%%%%%%%%%%%%%%%%%%%%%%%%%%%%%%%%%%%%%%

This research was supported in part by an NSF-BMBF (USA-Germany) CRCNS grant (NSF \#2112862 and BMBF \#STAXS).
The simulations were performed on the Pegasus, Foundry, and Mill clusters at Missouri S\&T.

%%%%%%%%%%%%%%%%%%%%%%%%%%%%%%%%%%%%%%%%%%%%%%%%%%%%%%%%%%%%%%%%%%%%%%%%%%%%%%%
\section*{Data availability}
%%%%%%%%%%%%%%%%%%%%%%%%%%%%%%%%%%%%%%%%%%%%%%%%%%%%%%%%%%%%%%%%%%%%%%%%%%%%%%%

The data that support the findings of this article are openly available \cite{House25}.

%%%%%%%%%%%%%%%%%%%%%%%%%%%%%%%%%%%%%%%%%%%%%%%%%%%%%%%%%%%%%%%%%%%%%%%%
\appendix
%%%%%%%%%%%%%%%%%%%%%%%%%%%%%%%%%%%%%%%%%%%%%%%%%%%%%%%%%%%%%%%%%%%%%%%%
\section{Interaction with finite decay time}
\label{app:decay}
%%%%%%%%%%%%%%%%%%%%%%%%%%%%%%%%%%%%%%%%%%%%%%%%%%%%%%%%%%%%%%%%%%%%%%%%%%%%

In this Appendix, we consider a generalization of FBM with mean-density interaction to cases in which the influence of the density at previous times decays with time. This can be modeled by introducing a memory kernel into the definition of the integrated density. Specifically, we replace the total integrated density $P_\mathrm{tot}(x, t_n) = N P_\mathrm{int}(x, t_n)$  defined in Eq.\ (\ref{eq:Pc_discrete}) with
\begin{eqnarray}
P_\mathrm{tot}(x, t_n) &=&  \sum_{m=1}^n   e^{-(t_n-t_m)/\tau_d} \, \sum_{j=1}^N  \delta [x- x^{(j)}_m] ~, \quad
\label{eq:Pc_discrete_decay}
\end{eqnarray}
where $\tau_d$ is the decay time. This implies that the effect of past densities on the current force decays exponentially with time. The theory  developed in the main part of the paper is recovered for $\tau_d \to \infty$.
It is clear that the decaying memory kernel  weakens the forces compared to the $\tau_d=\infty$ case. Thus, the results  for $\alpha> 4/3$ will not be affected qualitatively by the memory kernel. For long times the motion will remain to be noise-dominated with $\langle x^2 \rangle \sim t^{\alpha}$. In contrast, the long-time behavior for $\alpha < 4/3$ (which is force-dominated for $\tau_d =\infty$) is expected to  change for finite $\tau_d$.

To analyze this, let us first discuss how the scaling theory of Sec.\ \ref{sec:scaling} is modified by this change. We need to distinguish times $t \ll \tau_d$
and $t \gg \tau_d$.  For short times, $t \ll \tau_d$, the exponential decay factor in Eq.\ (\ref{eq:Pc_discrete_decay}) can be neglected.
The resulting theory is thus identical to the theory of Sec.\  \ref{sec:scaling}. In one dimension and for $\alpha < 4/3$, the mean-squared displacement is thus expected to behave as $\langle x^2 \rangle \sim t^{4/3}$ for $t \ll \tau_d$ (after the initial crossover at $\tau_x$ to interaction-dominated behavior discussed in Secs.\ \ref{sec:results-msd} and \ref{sec:scaling}).

For long times, $t \gg \tau_d$ the scaling theory changes. Specifically, the scaling ansatz (\ref{eq:scaling}) needs to be replaced by
 \begin{equation}
P_\mathrm{int}(x,t) = \frac {\tau_d} {b(t)} Y\left [\frac{x} {b(t)} \right]
\label{eq:scaling_decay}
\end{equation}
because the space integral (normalization) of $P_\mathrm{int}(x,t)$ is proportional to $\tau_d$ rather than $t$. Following the steps outlined in Sec.\ \ref{sec:scaling} yields that the time-dependence of typical force varies as $t^{-2 \delta}$ with time. If these forces dominate over the noise, they produce a typical displacement
$x_\mathrm{typ}(t) \sim t^{1-2\delta}$. Self consistency requires that $x_\mathrm{typ}(t)$ behaves as $b(t)$. This implies
$\delta = 1 -2\delta$ or $\delta=1/3$.

We therefore conclude that a new critical value of $\alpha$ emerges in the long-time regime $t \gg \tau_d.$ For $\alpha < 2/3$,
the long-time behavior remains force-dominated even after $\tau_d$, and the mean-squared displacement increases as $\langle x^2 \rangle \sim t^{2/3}$. For $\alpha>2/3$, the noise dominates over the forces, leading to   $\langle x^2 \rangle \sim t^{\alpha}$.  The expected behaviors of the mean-squared displacement in the different regimes of one-dimensional FBM with mean-density interaction in the presence of a finite decay time are summarized in Table \ref{tab:regimes}.
\begin{table}
\renewcommand*{\arraystretch}{2}
\begin{tabular*}{\columnwidth}{c @{\extracolsep{\fill}} ccc}
		\hline\hline
		$\alpha$                                                &  $t<\tau_x$ & $\tau_x < t < \tau_d$ & $t > \tau_d$  \\ \hline
		$ \alpha > 4/3$        &      $t^\alpha$     &  $t^\alpha$  &  $t^\alpha$  \\
		$2/3 < \alpha < 4/3$& $t^\alpha$  & $t^{4/3}$  &  $t^\alpha$   \\
		$\alpha < 2/3$  &  $t^\alpha$ &  $t^{4/3}$        &  $t^{2/3}$   \\
		\hline\hline
	\end{tabular*}
	\caption{Behavior of the mean-squared displacement $\langle x^2 \rangle$ of one-dimensional FBM with mean-density interaction in the presence of a finite decay time $\tau_d$. $\tau_x$ is the initial crossover time to interaction-dominated behavior discussed in Secs.\  \ref{sec:results-msd} and  \ref{sec:scaling}. }
	\label{tab:regimes}
\end{table}

To test these predictions of the scaling theory, we have performed computer simulations of one-dimensional FBM with mean-density interaction for several values of the decay time $\tau_d$. We have focused on the regimes $\alpha < 2/3$ and $2/3 < \alpha < 4/3$
because no change in behavior is expected for $\alpha > 4/3$ (where the motion is already noise-dominated in the absence of the decaying memory kernel).

\begin{figure}
\includegraphics[width=\columnwidth]{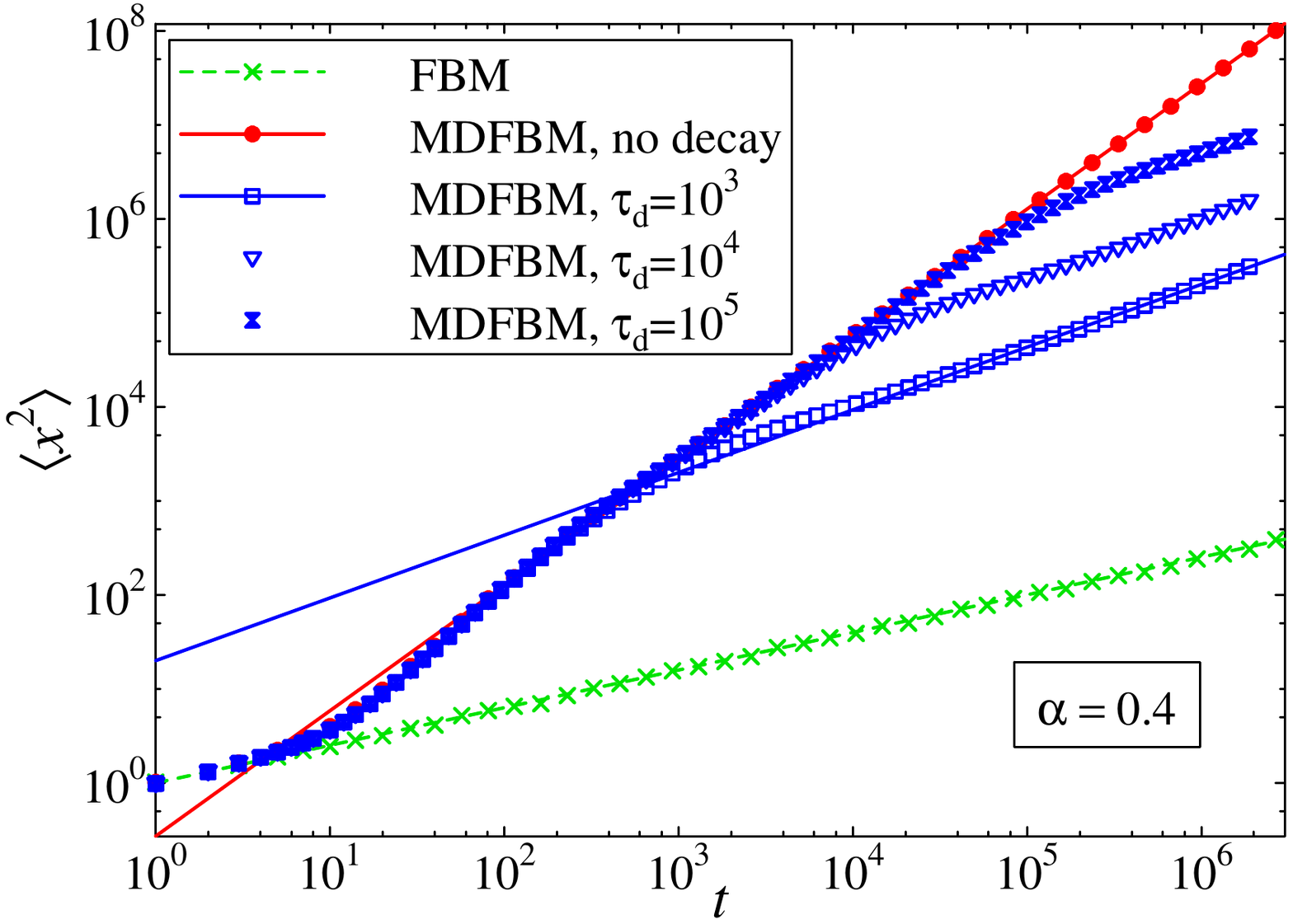}
\includegraphics[width=\columnwidth]{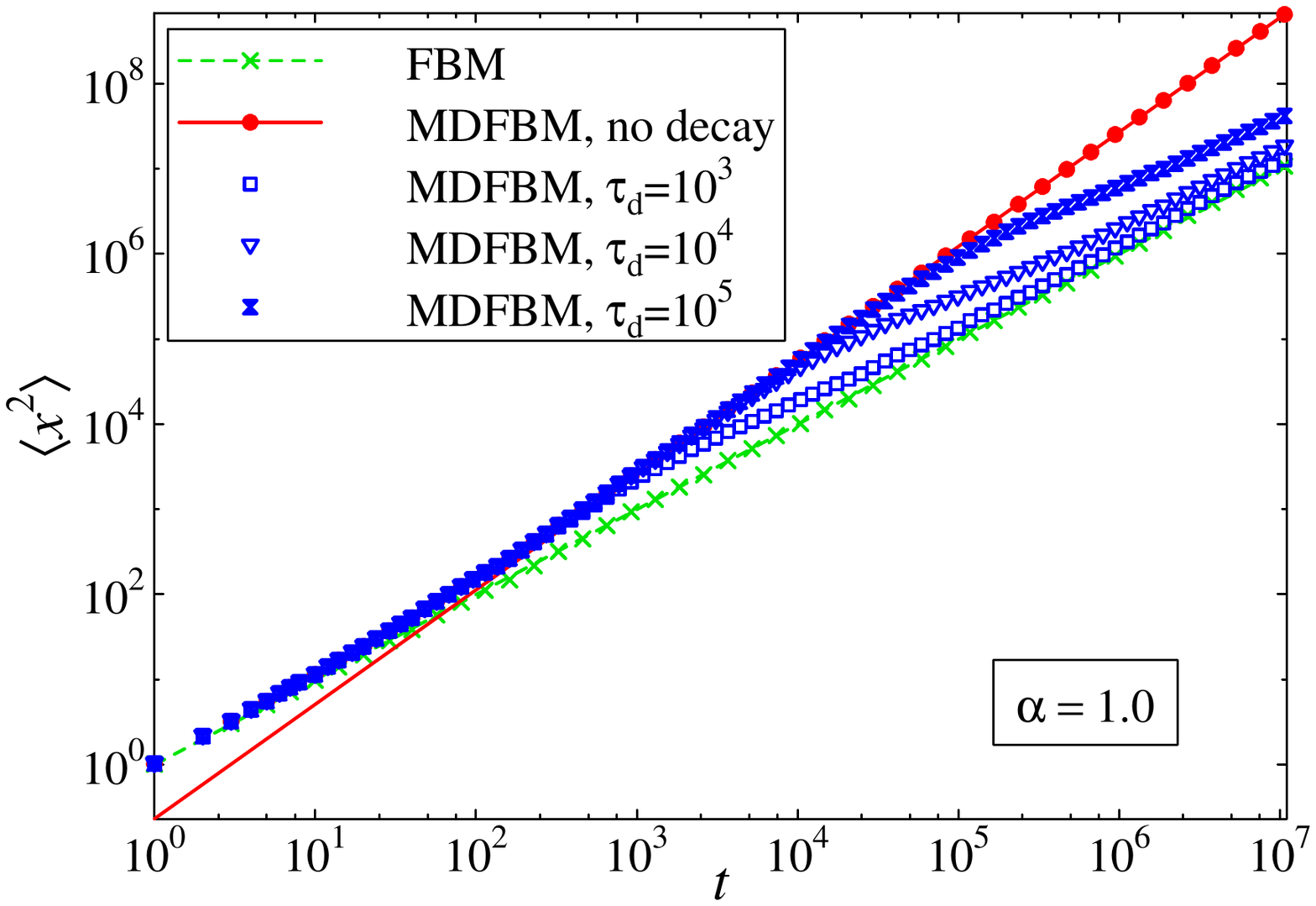}
\caption{Mean-squared displacement $\langle x^2 \rangle$ of FBM with mean-density interaction vs.\ time $t$ for $\alpha=0.4$ and 1.0 with interaction strength $A=1/5$ for several values of the decay time $\tau_d$. The data for noninteracting FBM are shown for comparison.  The data are averages over 16 ensembles of 1024 random walkers each.  The solid and dashed lines are power-law fits, for details see text.}
\label{fig:msd_decay}
\end{figure}

Figure \ref{fig:msd_decay} presents the mean-squared displacements for $\alpha=0.4$ and 1.0 for several values of the decay time $\tau_d$.
 To ensure that the three time regimes predicted by the scaling theory (see Table \ref{tab:regimes}) are well separated, we have chosen a comparatively strong interaction, $A=1/5$, for these simulations. This reduces the crossover time $\tau_x$ from the initial FBM regime
 to the interaction-dominated regime. The data for $\alpha=0.4$ (shown in the  upper panel of the figure) clearly feature the three predicted regimes. Initially, for time $t <\tau_x \approx 10$, the mean-squared displacement follows the FBM relation  $\langle x^2 \rangle \sim t^{0.4}$. At $\tau_x$, it crosses over to the interaction-dominated behavior  $\langle x^2 \rangle \sim t^{4/3}$ established in the main part of the paper. After the second crossover at $t=\tau_d$, the long-time behavior reaches the second interaction-dominated regime with  $\langle x^2 \rangle \sim t^{2/3}$. This can be seen clearly in the figure where the solid line for $\tau_d=10^3$ represents a power-law fit with the exponent $2/3$.

 The data for $\alpha=1.0$ (shown in the lower panel of Fig.\ \ref{fig:msd_decay}) feature similar behavior. For $t < \tau_x \approx 200$,
 the mean-squared displacement is governed by the FBM relation  $\langle x^2 \rangle \sim t^{1.0}$. Force-dominated behavior,    $\langle x^2 \rangle \sim t^{4/3}$, is observed between $\tau_x$ and $\tau_d$. For times $t > \tau_d$, the behavior becomes noise-dominated again and crosses back over  to $\langle x^2 \rangle \sim t^{1.0}$. This differs for the case of $\alpha=0.4$ discussed above where the long-time behavior remains interaction-dominated.

 The computer simulations thus fully confirm the predictions of the scaling theory for both $\alpha=0.4$ and $\alpha=1.0$.

\smallskip

%%%%%%%%%%%%%%%%%%%%%%%%%%%%%%%%%%%%%%%%%%%%%%%%%%%%%%%%%%%%%%%%%%%%%%%%
\section{Integrated density of FBM}
\label{app:idensity}
%%%%%%%%%%%%%%%%%%%%%%%%%%%%%%%%%%%%%%%%%%%%%%%%%%%%%%%%%%%%%%%%%%%%%%%%%%%%

In this Appendix, we sketch the derivation of  expression (\ref{eq:FBM_Pint}) for the integrated density of (non-interacting) FBM.
Equations (\ref{eq:Pint_discrete2}) and (\ref{eq:Pi}) imply that the integrated density simply is a sum over the (instantaneous) probability  densities,
\begin{equation}
P_\mathrm{int}(x, t_n) = \sum_{m=1}^n P(x,t_m) ~.
\label{eq:Pint_summation}
\end{equation}
As we are interested in the continuum limit $t \gg \epsilon=1$ (or $n \gg 1$), the sum can be replaced by an integral which reads
\begin{eqnarray}
P_\mathrm{int}(x, t) &=& \int_0^t d\tau P(x,\tau) \nonumber\\
&=& \int_0^t d\tau  \frac{1}{\sqrt{2\pi \sigma^2 \tau^\alpha}} \exp{ \left( -\frac{x^2}{2 \sigma^2 \tau^\alpha} \right) } ~.
\label{eq:Pint_integral}
\end{eqnarray}
Substituting $z=x^2/(2\sigma^2\tau^\alpha)$ leads to
\begin{equation}
P_\mathrm{int}(x, t) = \frac{|x|^{2/\alpha-1} }{\alpha \pi^{1/2}(2\sigma^2)^{1/\alpha}}
         \int_{x^2/(2\sigma^2 t^\alpha)}^\infty  dz\, z^{-1/\alpha-1/2} e^{-z}~.
\label{eq:Pint_subs}
\end{equation}
The integral over $z$ yields the upper incomplete $\Gamma$ function, which concludes the derivation of Eq.\ (\ref{eq:FBM_Pint}).
It is worth emphasizing that expression (\ref{eq:FBM_Pint}) fulfills the scaling form (\ref{eq:scaling}) with
$b(t)= x_\mathrm{rms}(t)=\sigma t^{\alpha/2}$. This can be seen explicitly by rewriting (\ref{eq:FBM_Pint}) as
\begin{eqnarray}
P_\mathrm{int}(x, t) &=& \frac t {\sigma t^{\alpha/2}} \, Y\left( \frac{x}{\sigma t^{\alpha/2}}\right) ~, \\
Y(y) &=& \frac{1}{\alpha \pi^{1/2}2^{1/\alpha}}\,   |y|^{2/\alpha-1}   \Gamma\left( \frac 1 2 - \frac 1 \alpha, \frac {y^2} 2\right)~.~
\label{eq:FBM_Pint_scaling}
\end{eqnarray}

%%%%%%%%%%%%%%%%%%%%%%%%%%%%%%%%%%%%%%%%%%%%
% BIBLIOGRAPHY
%%%%%%%%%%%%%%%%%%%%%%%%%%%%%%%%%%%%%%%%%%%%
%\bibliographystyle{apsrev4-2}
\bibliography{../../00Bibtex/rareregions}

\end{document}